# Wide-angle spectrally selective absorbers and thermal emitters based on inverse opals

Alireza Shahsafi[1], Graham Joe[1], Soeren Brandt[2], Anna V. Shneidman[2], Nicholas Stanisic[1], Yuzhe Xiao[1], Raymond Wambold[1], Zhaoning Yu[1,3], Jad Salman[1], Joanna Aizenberg[2,4], Mikhail A. Kats[1,3,5]

[1] Department of Electrical and Computer Engineering, University of Wisconsin-Madison

[2] John A. Paulson School of Engineering and Applied Sciences, Harvard University

[3] Department of Physics, University of Wisconsin-Madison

[4] Department of Chemistry and Chemical Biology, Harvard University

[5] Department of Materials Science and Engineering, University of Wisconsin-Madison

Abstract

Engineered optical absorbers are of substantial interest for applications ranging from stray light reduction to energy conversion. We demonstrate a large-area (centimeter-scale) metamaterial that features near-unity frequency-selective absorption in the mid-infrared wavelength range. The metamaterial comprises a self-assembled porous structure known as an inverse opal, here made of silica. The structure's large volume fraction of voids, together with the vibrational resonances of silica in the mid-infrared spectral range, reduce the metamaterial's refractive index to close to that of air and introduce considerable optical absorption. As a result, the frequency-selective structure efficiently absorbs incident light of both polarizations even at very oblique incidence angles. The absorber remains stable at high temperatures (measured up to ~900 °C), enabling its operation as a frequency-selective thermal emitter. The excellent performance of this absorber/emitter and ease of fabrication make it a promising surface coating for passive radiative cooling, laser safety, and other large-area applications.

Electromagnetic-wave absorption based on metamaterials has been a rapidly growing motif in different research fields, including energy conversion [1] and thermoregulation [2]. At optical frequencies, engineered absorbers are used for stray light reduction in the visible range [3], [4], refractive-index sensing in the near-infrared range [5]–[7], and thermal emitters [8]–[11] in the mid-infrared range. To achieve high absorption, the reflection, transmission, and scattering must all be minimized; this can be realized by an absorbing structure or material that is impedance-matched to the incident medium [12].

Since the wavelength of optical-frequency electromagnetic radiation is on the order of hundreds to thousands of nanometers, typical engineered optical absorbers are comprised of nanostructured materials [13], [14]. At the same time, many applications require inexpensive large-area absorbers [15], [16], motivating designs that can be fabricated without the use of top-down lithography. Examples of such lithography-free absorbers include carbon nanotube (CNT) forests [3], [17], and lossy low-index films like polydimethylsiloxane (PDMS) [18].

Here, we demonstrate a wide-angle large-area absorber for mid-infrared light with polarization-independent spectral selectivity, realized using a metamaterial based on inverse opals (IOs), which are highly porous structures obtained through assembly of colloids that serve as a sacrificial structuring agent for a background matrix material (here we use sol-gel silica; more information about the synthesis can be found in *Sup. Info. 1*) [19]. In IOs, the size and arrangement of the pores and the material composition (background matrix, as well as dopants and other inclusions) can all be engineered, creating a large design space. The pores are also interconnected and can be infiltrated with fluids [20], which may be relevant for dynamic tunability, sensing, or catalytic applications [21]–[25].

The diameter of the templating colloidal spheres used for IOs is typically on the order of a few hundred nanometers. After selective removal of the colloids, the corresponding volume comprises an interconnected array of air pockets while the surrounding volume is the glassy matrix material. The resulting composite material thus has an effective refractive index close to that of air (in particular, when using silica as the matrix material), as has been observed at visible frequencies [26]. In the mid-infrared range, the vibrational resonances of silica [27] introduce substantial optical loss, resulting in an absorbing metamaterial that is well matched to the refractive index of air, thus minimizing reflections at the interface between air and the IO film.

A scanning electron microscope (SEM) image of an inverse opal film with void size of 250 nm is presented in Fig. 1(a), showing the highly interconnected porous structure. The high degree of ordering (which can be disrupted through the addition of salts in the assembly process) is unimportant for the functionality at mid-infrared wavelengths; however, it is convenient for the modeling of the effective refractive index, since the volume fraction of background matrix and voids is well known for a face-centered cubic (fcc) lattice, 26% and 74%, respectively [19].We performed infrared variable-angle spectroscopic ellipsometry (J.A. Woollam IR-VASE Mark II) on a 2-µm-thick IO film on silicon and extracted its optical properties (Fig. 1 (b), (c)). To convert the measured ellipsometric parameters, $\Psi$ and $\Delta$ [28] (see *Sup. Info. 2* for raw data), into the effective refractive index and absorption coefficient, we modeled the IO thin film using the Bruggeman effective medium approximation (EMA, see *Supp. Info. 2*) [29]. We first focused on 1.8 to 3 $\mu m$, a spectral range in which SiO2 is nearly dispersionless and its refractive index is well known [30], and fit to the fraction of air present in the film (i.e. the film porosity). The resulting fitted porosity is ~76%, which is very close to the theoretical value of the porosity based on the fcc structure of the opal assembly ($\sqrt{2}\pi/6 = 74\%$) [19]. Fixing the porosity at this value, we then fit the ellipsometric measurements in the wavelength range of 6 to 12 µm, in which silica has vibrational resonances. The resulting Kramers–Kronig-consistent fitted effective complex refractive index of the inverse opal film, $n_{IO} + i\kappa_{IO}$, along with the fitted values assuming no porosity are shown in Fig. 1(b, c); details about the fitting parameters are provided in *Sup. Info. 2*. The void size of the IOs (~250 nm) is significantly smaller than the wavelength in the mid-infrared range, ensuring the validity of the effective-medium assumption [31]. For comparison, the optical properties of an amorphous SiO2 thin film from literature [30] are also displayed in the same figure. The IO film has $n_{IO}$ very close to one ($n_{air} = 1$), and a significantly reduced but non-zero $\kappa_{IO}$ compared to the bulk amorphous SiO2.

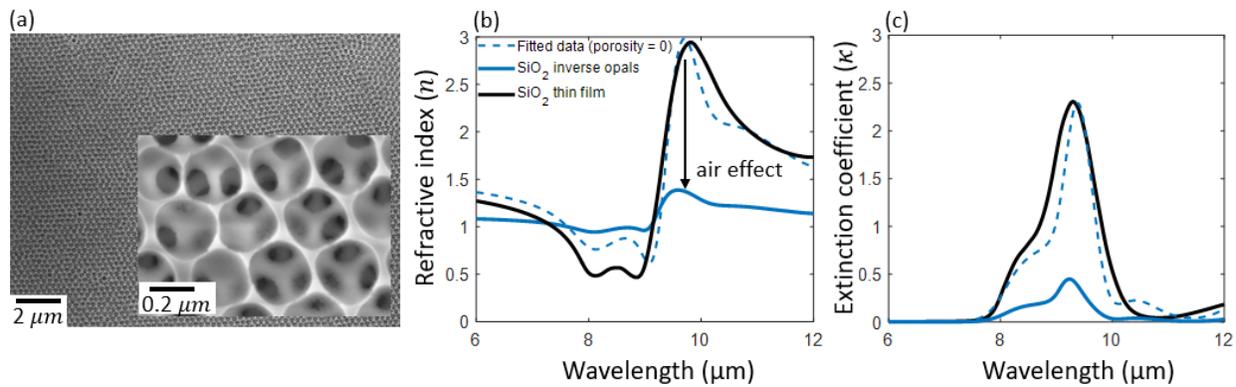

**Figure 1) (a)** Scanning electron microscope (SEM) image of the inverse opals (IO) used in this study, showing the nano-porous interconnected structure with a void diameter of 250 nm. The inset SEM image of the same sample, magnified 10×, shows the high porosity and interconnectivity the IO. **(b)** Real ($n$) and **(c)** imaginary ($\kappa$) parts of the complex refractive index of the inverse opal film (solid blue lines) along with the fitted values assuming no porosity (dashed blue lines), and bulk sputtered amorphous SiO2 (black lines) [19], for wavelengths close to the vibrational resonances of SiO2. Optical properties of inverse opals are measured via variable-angle ellipsometry and fitted using the Bruggeman effective-medium theory [21].

Using the extracted refractive indices in Fig. 1(b, c) and the transfer-matrix method [32], [33] (*See Sup. Info. 3*), we calculated the optical impedance, $Z_{IO}$, of 2- and 4-$\mu m$ thick IO film on a silicon substrate. Figure 2(a) shows the magnitude of the difference between $Z_{IO}$ and the free-space impedance, $Z_0$. We also calculated the impedance difference, $|Z_{ox} - Z_0|$, between $Z_0$ and a homogeneous silica film of the same thickness on the same silicon substrate for the two thicknesses of 2 and 4 $\mu m$ (Fig. 2(a), dashed lines). In a portion of the mid-infrared spectral range, the combination of the large volume fraction of air inclusions and the presence of strong vibrational resonances in silica results in a close match between the impedance of the IO structure and that of air; i.e., $|Z_{IO} - Z_0|$ approaches zero. We observed a large contrast between the $|Z_{ox} - Z_0|$ and $|Z_{IO} - Z_0|$, especially near the vibrational resonances.

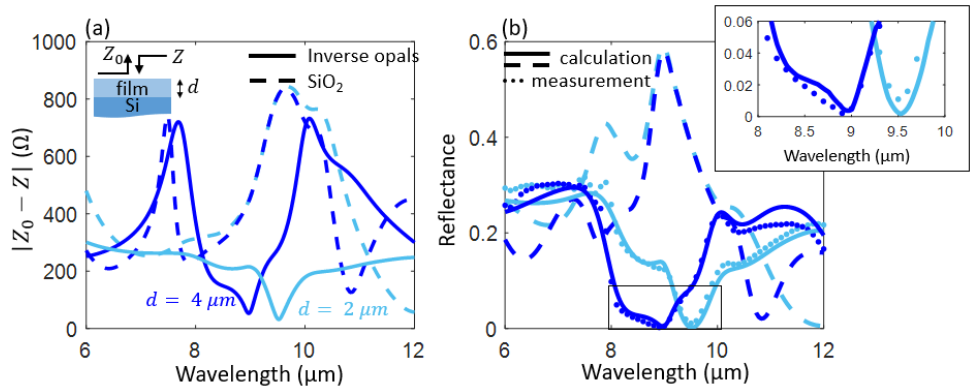

**Figure 2) (a)** The difference between the optical impedance of free space, $Z_0 = 377$ Ω, and the calculated impedance (using the transfer-matrix method), of a structure comprising a layer of IOs ($Z_{IO}$, solid lines) or homogeneous SiO$_2$ ($Z_{ox}$, dashed lines) at two different thicknesses (2 and 4 µm), on top of a semi-infinite silicon substrate. **(b)** Measured (dotted) and calculated (solid lines) reflectance of the IO films, as well as bulk SiO$_2$ (calculated, dashed lines), at near-normal incidence. The reflectance of the IO film is suppressed due to impedance matching with air for wavelengths of 8 to 10 µm. The inset zooms in on the region of lowest reflectance.

To confirm the absorption around the resonances, we performed reflectance measurements using Fourier transform spectroscopy (FTS). Figure 2(b) shows the calculated (solid lines) and measured (dotted lines) values for the reflectance of the IO films at two thicknesses (2 µm and 4 µm) on top of a doped silicon substrate. The calculation was performed via the transfer-matrix method, using the optical properties extracted from ellipsometry [Fig. 1(b-c)]. The measurement was made using a Bruker Vertex 70 coupled to a Hyperion 2000 infrared microscope, with a numerical aperture of 0.4. As expected, the reflectance of the IO film becomes very small in the spectral region where the impedance is nearly matched to that of free space. In the same wavelength range, the reflectance of thin-film SiO$_2$ is quite large due to the large impedance mismatch.

We subsequently analyzed the angle-dependent reflectance of two IO films of different thicknesses (2 and 4 µm, as before) on silicon, for both s- and p- polarizations (Fig. 3). For both thicknesses, the reflectance decreases to nearly zero close to the spectral position of the vibrational resonances. Note that changing the film thickness affects the bandwidth of the suppressed reflectance region. Increasing thickness results in more light absorption in the spectral region where the IO film has a lower absorption coefficient. The optical impedance of the structure at oblique incidence is provided in *Sup. Info. 3*.

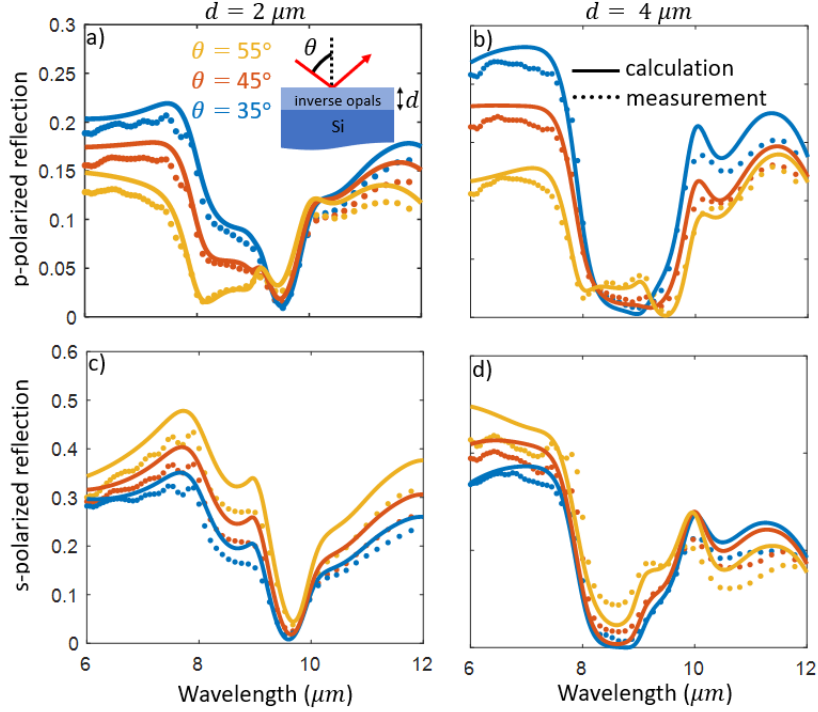

**Figure 3)** Calculated (solid line) and measured (dots) reflectance at oblique incidence (θ = 35°, 45°, and 55°) for p- and s-polarized light for two different thicknesses of the inverse opal films (d = 2 and 4 μm).

The selective absorptivity of the IO structures implies that they can serve as a selective thermal emitters; this is encoded in Kirchhoff's law of thermal radiation, which equates the absorptivity at every wavelength (and angle and polarization) to the object's emissivity, assuming equilibrium conditions [11], [16]. Kirchhoff's law is also often used in the characterization of thermal emission from non-scattering samples, because it is frequently easier to infer absorptivity from reflectance and transmittance measurements than to directly measure thermal emission [32], [33]. For the samples in this work, however, it was difficult to apply Kirchhoff's law directly to our measurements because the semi-transparent silicon substrate was single-side polished, and the scattering from the back side is difficult to quantify.

To obtain accurate measurements of the emissivity, we performed direct emission measurements using FTS, as previously described in refs. [34], [35]. We collected the thermal emission from the samples at 150 °C and normalized the values to the emission from a laboratory blackbody reference at the same temperature. A forest of vertically aligned CNTs (height = 0.5 mm) on a silicon wafer was used as the reference. We calibrated the emissivity of the CNT forest by comparing measured emission to the well-characterized thermal emission from flat wafers of fused silica and sapphire. Detailed information regarding this measurement is provided in *Sup. Info.* 4.

Using the direct thermal-emission technique, we measured the polarized oblique-angle emissivity by rotating the sample and placing a polarizer directly in front of it; this was done for s- and p-polarizations at three oblique angles [Fig. 4(a, b)]. The accuracy of angle- and polarization-resolved direct thermal-emission measurements were confirmed by performing similar measurements on flat wafers of fused silica and sapphire; see *Sup. Info. 5*.

For further confirmation of the measurements, we calculated the expected emissivity of this structure using Kirchhoff's law and the extracted optical properties of the IO films in Fig. 1(b-c). Two different emissivity spectra were calculated as presented in Fig. 4(a-b); one set of curves corresponds to the structure's emissivity assuming zero transmission through the Si substrate, (*i.e.*, absorptance $A = 1 - R$ when $T_{Si} = 0$; dotted lines), and the other assumes a lossless substrate (*i.e.*, $A_{Si} = 0$, $A = 1 - R - T$; dashed lines). For the spectral regions where the IO film has relatively low loss ($\lambda < 8$ μm or $> 10$ μm), these calculations provide upper and lower bounds on the actual emissivity.

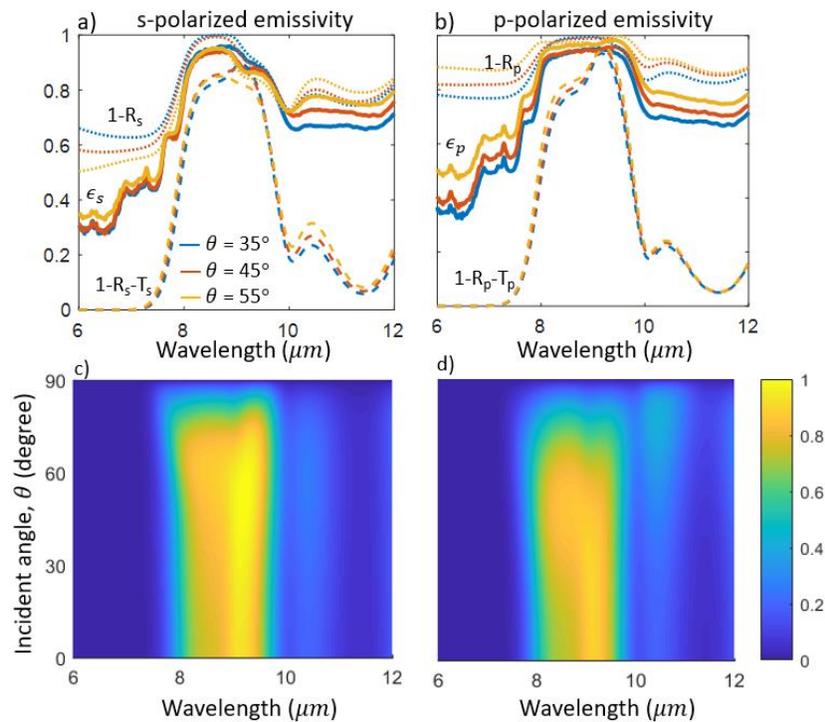

**Figure 4)** Experimental emissivity (solid lines) via direct-emission measurements at 150 °C for 4-μm-thick IO on silicon wafer at different incidence angles ($\theta$ = 0°, 35°, 45°, and 55°) for (**a**) s-polarized light and (**b**) p-polarized light, along with calculated upper (dotted lines) and lower bounds (dashed lines) on the emissivity in the spectral range where the IO film is low loss (λ < 8 μm or > 10 μm). In the spectral range where the IO film is low loss (λ <8 μm or >10 μm), we expect a substantial contribution to the emissivity from the lossy silicon substrate. Calculated absorption (1 – transmission – reflection) within the film is shown versus wavelength and incident angle for (**c**) s polarization and (**d**) p polarization.

The large-area near-unity absorption at oblique incidence can be readily visualized using long-wave infrared ("thermal") imaging. Infrared-camera images typically attribute a pixel brightness to the intensity of detected thermal radiation from that pixel, assuming a constant emissivity for all pixels of the image. In Fig. 5(a), we show such infrared images of a silica glass slide (labeled

"silica" in the image) next to the IO sample measured in Fig. 4. We used a FLIR A325sc camera with a bandwidth of 7.5 to 13.0 µm. To isolate the high-emissivity region of the IO identified in Fig. 4, we positioned a filter with a passband of 8.2 to 10.6 µm in front of the camera. The infrared images with and without the filter can be seen in the bottom and top rows of Fig. 5, respectively, with the images taken from the normal direction, as well as for oblique angles of 35°, 45°, and 55°. These images demonstrate the large and broad-angle spectral selectivity of our selective thermal emitter. The "lines" of high apparent temperature seen along the bottom edges of the silica samples are due to roughness, which creates a gradient of refractive index, and thus enhances emissivity, similar to the effect for other porous structures; see *Sup. Info.* 6. Note that since our FLIR infrared camera and the readout software have a fixed built-in algorithm to detect the temperature using measured thermal radiation in the 8 to 14 *µm* range, it is no longer possible to directly interpret the infrared image as a temperature map when the object is viewed through a bandpass filter. Instead, in Fig. 5 we report an "Apparent temperature", defined as the temperature reported by the camera software, without any adjustments made due to the presence of the filter.

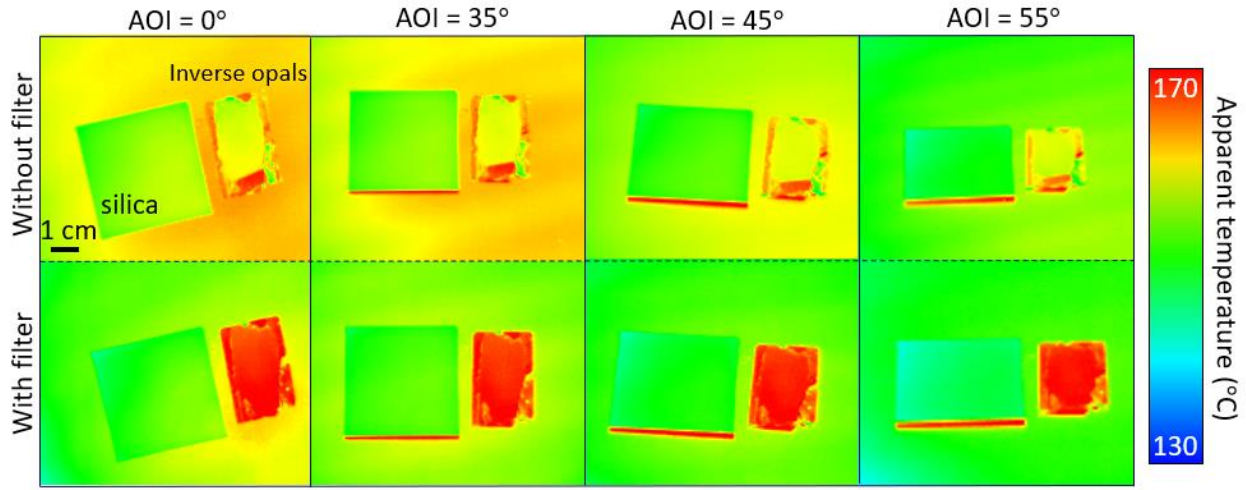

**Figure 5)** Mid-infrared images of our IO absorber and a reference silica glass slide at 150 °C, imaged at angles (AOI) of 0 to 55°, without (top) and with (bottom) a filter that selects the wavelength range corresponding to high absorptivity/emissivity of the sample (8.2 to 10.6 µm). The close-to-unity emissivity of the IO absorber results in a large apparent temperature.

In conclusion, we demonstrated a large-area mid-infrared absorber/emitter based on a self-assembled silica inverse-opal (IO) metamaterial with a thickness less than half of the free-space wavelength. The device shows little angular and polarization dependence across its working wavelength of 8 to 10 µm, maintains a greater than 80% absorption for incidence angles as large as 80°, and is stable up to at least ~900 °C. The broad-angle absorption is a consequence of impedance matching to the near-unity refractive index of the IO structure with considerable optical losses, resulting from the combination of mid-infrared vibrational resonance of the silica comprising the opal matrix and the large volume fraction of air inclusions. By utilizing alternative matrix materials with a different set of vibrational resonances and/or by depositing materials within the IO voids, the self-assembly approach can enable large-area wide-angle absorbers and thermal emitters across the infrared range. Further functionality may be realized with full or partial fluid infiltration through the structure to achieve dynamic tunability, or for sensing or photo-thermal catalysis. Various applications may also benefit from the high degree of order achievable with IOs, including simultaneous optical effects at visible wavelengths.

## Acknowledgements

MK acknowledges funding from the National Science Foundation (NSF, ECCS-1750341) and the Department of Energy (DOE, DE-NE0000743). JA acknowledges funding from the U. S. Army Research Office (ARO, W911NF-17-1-0351). SB and AVS are supported in part by the Kavli Institute for Bionano Science and Technology at Harvard. This work was performed in part at the Center for Nanoscale Systems (CNS), a member of the National Nanotechnology Coordinated Infrastructure Network (NNCI), which is supported by the National Science Foundation under NSF ECCS award no. 1541959. This work was also performed in part at the Wisconsin Centers for Nanoscale Technology (WCNT), core facilities at UW-Madison. We also acknowledge use of facilities and instrumentation supported by NSF through the UW-Madison Materials Research Science and Engineering Center (DMR-1720415).

# Supplementary Information:

## Wide-angle spectrally selective absorbers and thermal emitters based on inverse opals

Alireza Shahsafi[1], Graham Joe[1], Soeren Brandt[2], Anna V. Shneidman[2], Nicholas Stanisic[1], Yuzhe Xiao[1], Raymond Wambold[1], Zhaoning Yu[1,3], Jad Salman[1], Joanna Aizenberg[2,4], Mikhail A. Kats[1,3,5]

[1] Department of Electrical and Computer Engineering, University of Wisconsin-Madison

[2] John A. Paulson School of Engineering and Applied Sciences, Harvard University

[3] Department of Physics, University of Wisconsin-Madison

[4] Department of Chemistry and Chemical Biology, Harvard University

[5] Department of Materials Science and Engineering, University of Wisconsin-Madison

## 1. Inverse opal synthesis

We synthesized inverse-opal (IO) films using an evaporative co-assembly method based on a procedure by Hatton *et al*. [S1]. The templating colloids were monodisperse polystyrene particles with diameters ranging from 120 to 450 nm. Aqueous solutions of the colloids were synthesized by surfactant-free emulsion polymerization [S2]. For co-assembly, we diluted these solutions to 0.1 v% particle content. We prepared the sol-gel precursor for the silica background matrix from a solution of 1.5 mL ethanol (100%, VWR), 1mL tetraethyl orthosilicate (TEOS, 98% Sigma Aldrich), and 1mL 0.1 M hydrochloric acid (HCl, Sigma Aldrich) solution, which was stirred at room temperature for one hour before use. The final co-assembly solution was prepared by adding 60 $\mu L$ of the TEOS sol-gel solution to 12 mL of the colloid solution. For co-assembly, we used flat silicon substrates (approximately 1 by 4 cm). We prepared them by cleaning in piranha (3:1 sulfuric acid (Sigma): hydrogen peroxide (VWR)) for 30 min, washing in water and ethanol, and drying with compressed air. We suspended the substrates vertically in the co-assembly solution, which was then allowed to evaporate over 2 to 3 days at 65 °C. After the water fully evaporated, we produced the IO films by removing the colloidal template at 500 °C in air over 12 hours (5 h ramp time, 2 h at 500 °C, 5 h ramp back to room temperature).

## 2. Ellipsometry measurement

Infrared spectroscopic ellipsometry measurements were performed on the synthesized IO films. Measurements were performed at three different angles of incidence: 35, 45, and 55°. Compared to infrared wavelengths, the feature size of IO films is very small, allowing the film to be modeled as a single effective medium. The films were modelled using an isotropic Bruggeman effective medium approximation ($p \frac{\epsilon_{eff} - \epsilon_{ox}}{2\epsilon_{eff} + \epsilon_{ox}} = (p-1) \frac{\epsilon_{eff} - 1}{2\epsilon_{eff} + 1}$, $p$ is the volume fraction of the silica with permittivity of $\epsilon_{ox}$) in which the composition of air and silica was tunable. The porosity used in this model is measured through near infrared ellipsometry. These parameters were then used in the longer wavelength infrared model of the IO film to fit a Kramers-Kronig-consistent oscillator model to the measured ellipsometry parameters Ψ and Δ. Experimental results and the corresponding fitted values are shown with dashed and solid lines in the Fig. S1, respectively.

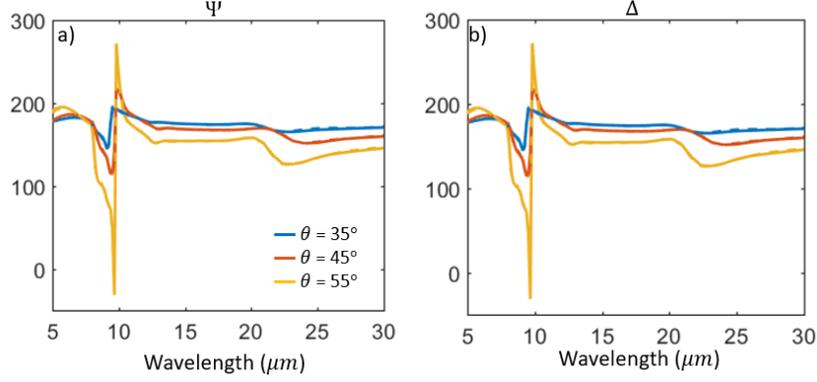

**Figure S1)** Fitted (dashed) and measured (solid) ellipsometry variables for inverse opals at three different incidence angles, $\theta$, for (a) $\Psi$ and (b) $\Delta$.

## 3. Impedance calculation

To investigate the zero-reflection dip of the metamaterial (Fig. 4), we analyzed the impedance mismatch between our structure ($Z$) and free space ($Z_0 = 377\ \Omega$) (Fig. S2). Impedance for the structure can be calculated through:

$$Z = \frac{1+r}{1-r} Z_0,$$

Where $r$ is the complex Fresnel coefficient of reflection from the structure. The magnitude of the impedance mismatch, $|Z - Z_0|$, is presented in Fig. 2(a) for two different thicknesses. Here we plot the impedance mismatch as a function of wavelength and angle a for 4-$\mu m$-thick IO film on a lossless silicon substrate for both s- and p- polarization.

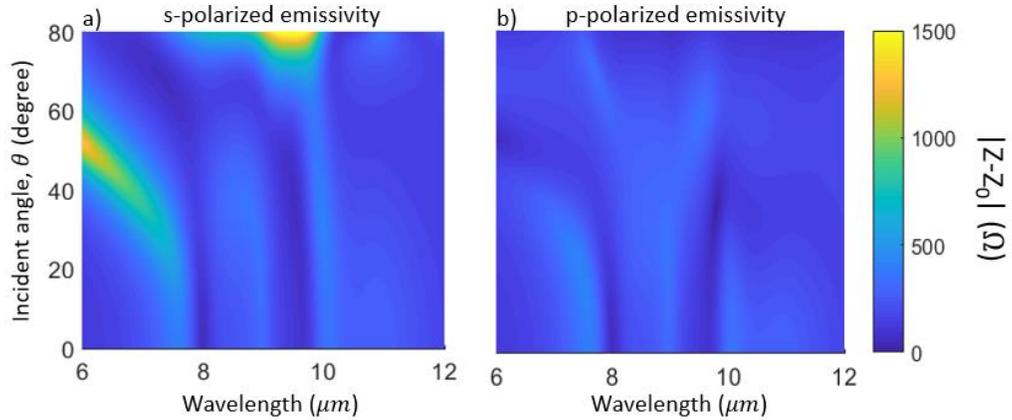

**Figure S2)** Spectral and angular $|Z-Z_0|$ for (a) s polarization, and (b) p polarization.

## 4. Optical paths of emission measurement and emissivity for normal incidence

To confirm the accuracy of the direct emission measurement to measures the emissivity of our absorber, we performed thermal emission measurements at normal incidence in two different ways using a Fourier transform spectrometer (Bruker Vertex 70 FTIR). The first way (used in the main text) had the sample in the sample compartment of the FTS, where we also mounted an angle-variable stage to enable the angle-dependent measurements described in the main text (Fig. S3 (a)). The emitted light was detected using a mercury cadmium telluride (MCT) detector in a Bruker Hyperion 2000 optical microscope connected to the output port on the right in Fig. S3(a). The second method uses the same FTS, but with the sample at the situated in the microscope, and the beam traversing the opposite direction through the FTS, into a different MCT detector on the other side [Fig. S3 (b)]. We measured the normal-direction emissivity of the 4-µm-thick IO sample on silicon and found the same results, performing the measurement in both directions [Fig. S3(c)].

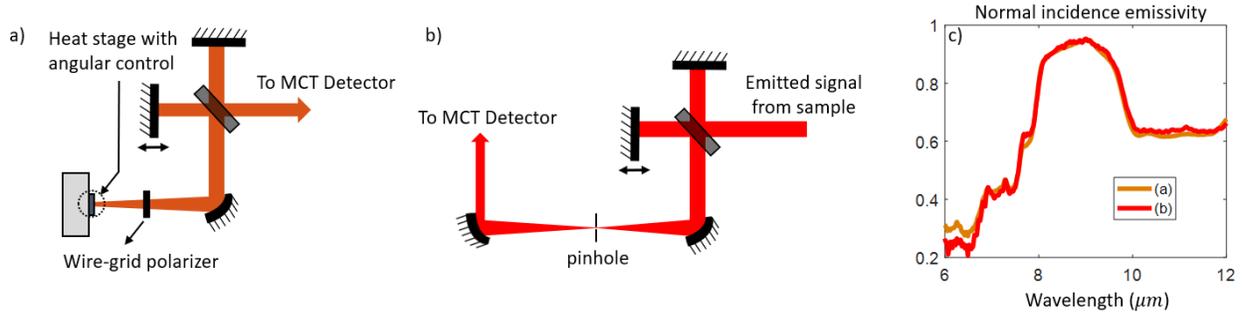

**Figure S3)** Diagrams that outline the two different paths for emission measurements in the Fourier-transform spectrometer (FTS) setup. (a) In this orientation, the emission signal comes from the custom-built angular stage holding the heater and the sample, while in (b) the emission signal comes from an infrared microscope setup to a different MCT detector, after the light passes through the interferometer. (c) Measured emissivity for the two different paths of previous figure.

To confirm the accuracy of angular emission measurement, we compare results from direct emission measurements with angular reflectance measurements. Figures S4(b-c) show the ratio between the emissivity of silica and sapphire for three different incident angles (35, 45 and 55°) for both the s- and p-polarization, measured with two different techniques: direct emission (solid lines) and reflection (1-R, dashed lines) measurements. As seen in the plots, regardless of the method used to extract emissivity, in highly lossy spectral region for both silica and sapphire ($\lambda > 7 \ \mu m$), the emissivities values are similar, validating the direct emission measurement at oblique incidence. Note that in the shorter wavelength spectral range that sapphire becomes semi-transparent, these ratios begin to vary because to calculate emissivity based on Kirchhoff's law we need to incorporate the transmission through sapphire wafer, which we do not do here. For both reflection and direct-emission measurements, we found out that the maximum uncertainty is less than 5%. This error comes mainly from the imperfection of our laboratory blackbody reference for emission measurements and the gold film reference for reflectance measurements. Also, for direct-emission measurements that do not use a microscope, accurate focusing is challenging and results in errors of up to 2% in the raw measured signal.

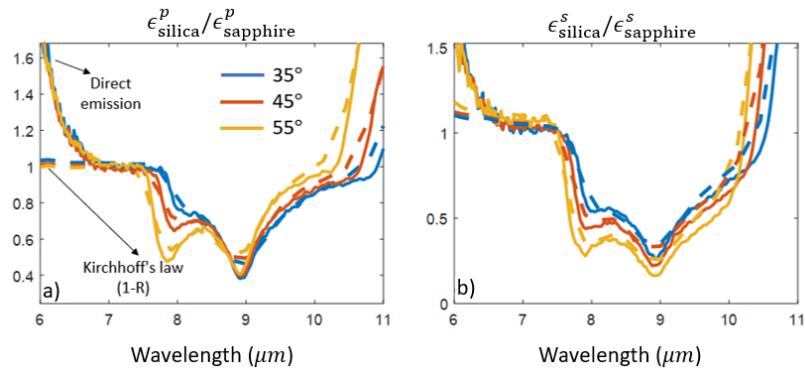

Figure S4) (a-b) Ratio of silica emissivity to sapphire emissivity at three different angles, resulted from direct emission measurement (solid lines) and oblique reflection measurement (dashed lines) and the application of Kirchhoff's law for (s) p-polarized, and (b) s-polarized light.

## 5. Emission measurement for semi-transparent substrate and oblique incidence

As noted in the main manuscript, the substrate of the IO sample is semi-transparent at mid-infrared wavelengths due to optical losses. Here, in Fig. S4(a), we show that the emission from the substrate (with no inverse opals) normalized to that of blackbody is significant. This explains the considerable emission in Fig. 4(a-b) outside of 8 to 10 μm.

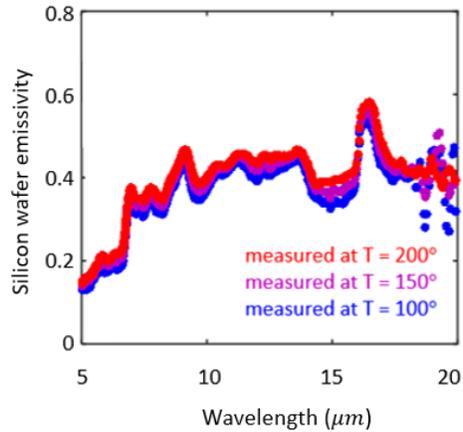

**Figure S5)** (a) Emission contribution of the silicon substrate alone.

## 6. Infrared imaging

We captured infrared images of the sample to illustrate its large absorption at oblique incidence angles over wide areas. Since our thermal camera's bandwidth (7.5 to 13 μm) differed from the working spectral range of our absorber (8 to 10 μm), we used a bandpass infrared filter in front of the camera to restrict the incoming infrared radiation to the absorber's spectral range (Fig. 5 top vs. bottom). The transmission spectrum of the filter used in this setup is provided in Fig. S6.

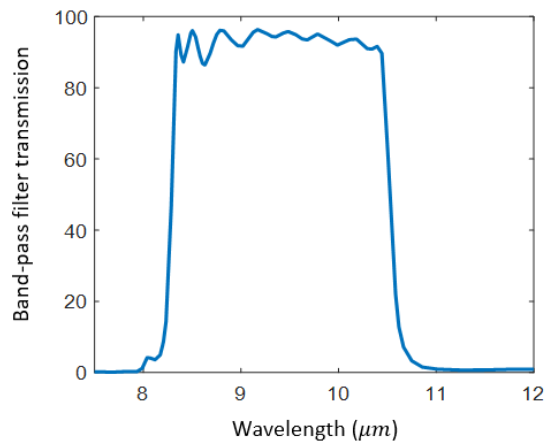

**Figure S6)** Spectral transmission of the employed 8.2 to 10.6 $\mu m$ bandpass filter.

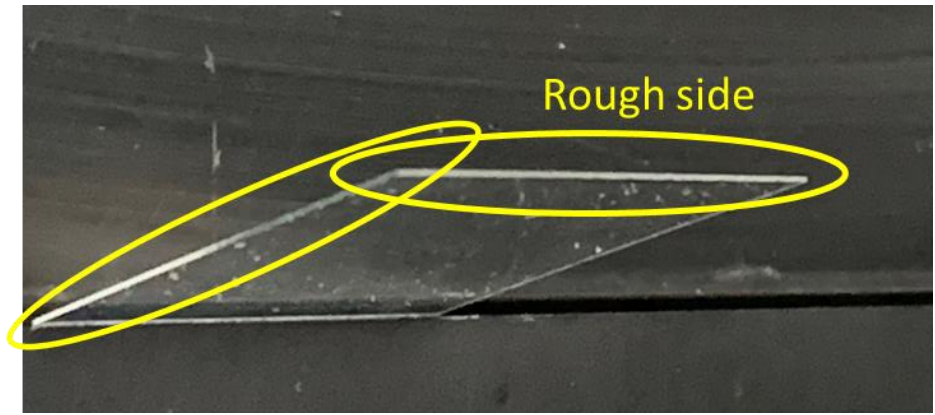

**Figure S7)** Image of rough edges of the silica wafer shown in Fig. 5

The sides of the silica wafer are rough because these are not high-quality cleaved samples (Fig. S7). This results in a gradient-index behavior on the edges, better matching the impedance between silica and air, suppressing the reflection similar to other porous structures [S3]. Thus, the edges are more emissive than the flat top surface of the silica wafers, as can be seen from the reddish regions around the edges in Fig. 5 in the main text.

S1) B. Hatton, L. Mishchenko, S. Davis, K. H. Sandhage, and J. Aizenberg. "Assembly of large-area, highly ordered, crack-free inverse opal films." *Proc. Natl. Acad. Sci.* vol. 107, no. 23, pp. 10354-10359, 2010.

S2) J. W. Goodwin, J. Hearn, C. C. Ho, and R. H. Ottewill "Studies on the preparation and characterization of monodisperse polystyrene lattices." Colloid. Polym. Sci. vol. 252, no. 6, pp. 464-471, 1974.

S3) X. Li, J. Gao, L. Xue, and Y. Han. "Porous polymer films with gradient‐refractive‐index structure for broadband and omnidirectional antireflection coatings." *Adv. Func. Mat.* vol. 20, no. 2, pp. 259-265, 2010.